\documentclass[12pt,a4paper]{article}
\usepackage{amsmath}
\usepackage{latexsym}
\makeatletter
\def\rddots{\mathinner{\mkern1mu\raise\p@%
    \vbox{\kern7\p@\hbox{.}}\mkern2mu%
    \raise4\p@\hbox{.}\mkern2mu\raise7\p@\hbox{.}\mkern1mu}}
\makeatother
\newcommand{\fukuso}{{\mathbf C}}
\newcommand{\futon}{{\bf N}}

\begin{document}

\title{\sl Reduced Dynamics from the Unitary Group to 
Some Flag Manifolds : Interacting Matrix Riccati Equations}
\author{
  Kazuyuki FUJII
  \thanks{E-mail address : fujii@yokohama-cu.ac.jp }\quad and\ 
  Hiroshi OIKE
  \thanks{E-mail address : oike@tea.ocn.ne.jp }\\
  ${}^{*}$Department of Mathematical Sciences\\
  Yokohama City University\\
  Yokohama, 236--0027\\
  Japan\\
  ${}^{\dagger}$Takado\ 85--5,\ Yamagata, 990--2464\\
  Japan\\
  }
\date{}
\maketitle
\begin{abstract}
  In this paper we treat the time evolution of unitary elements 
  in the $N$ level system and consider the reduced dynamics from 
  the unitary group $U(N)$ to flag manifolds of the second type 
  (in our terminology). Then we derive a set of differential 
  equations of matrix Riccati types interacting with one another  
  and present an important problem on a nonlinear superposition 
  formula that the Riccati equation satisfies.

  Our result is a natural generalization of the paper 
  {\bf Chaturvedi et al} (arXiv : 0706.0964 [quant-ph]).
\end{abstract}
%


%
%
%
%
\newpage

\section{Introduction}

In this paper we treat a finite quantum system (N--level system) 
and consider its unitary evolution in detail. 

The unitary evolution for a time--dependent Hamiltonian 
$H(t)$ (which is not so artificial) is given by
\[
i\hbar \frac{d}{dt}U(t)=H(t)U(t),\quad U(0)=E_{N}
\]
where $H(t)\in H(N,\fukuso)$, $U(t)\in U(N)$ and $E_{N}$ is the 
identity matrix. In the following we set $\hbar =1$ and write
\begin{equation}
i\dot{U}(t)=H(t)U(t),\quad U(0)=E_{N}
\end{equation}
for simplicity.

Now, we consider a reduction of symmetry. Namely, symmetry happens to 
reduce from the unitary group $U(N)$ to a subgroup $H$. Then the 
evolution on $U(N)$ ``split" into evolutions on the subgroup $H$ and 
corresponding homogeneous space $U(N)/H$. 

We are interested in the evolution on $U(N)/H$ with special $H$ and 
call this the reduced dynamics from $U(N)$ to $U(N)/H$ for short. 

Especially, for $H=U(m)\times U(n)$ ($m+n=N$) the homogeneous space 
is the Grassmann manifold and the matrix Riccati equation appears 
naturally, see \cite{6-persons} and \cite{UR}. 

Interesting enough, we meet the matrix Riccati equation(s) in several 
fields in Mathematics or Physics. In the special case $m=n=1$ we have 
usual (complex--valued) Riccati equation. This equation has a 
mysterious formula called a nonlinear superposition one, see the text 
in detail. 

In the paper we generalize the Grassmann manifold to some Flag manifold. 
Namely, for $H=U(l)\times U(m)\times U(n)$ ($l+m+n=N$) we call $U(N)/H$ 
the flag manifold of the second type (temporarily). Then a set of 
differential equations of matrix Riccati types interacting with 
one another are obtained. This is the main result and we also present a 
problem on a nonlinear superposition when $l=m=n=1$. 
See \cite{CGM} as a general introduction to this topic.

As a result it may be concluded that ``Riccati structure" appears 
naturally in the process of symmetry reduction from the unitary group 
to some subgroup.

\section{Reduced Dynamics on Grassmann Manifolds : Review}

We set $m+n=N\ (m, n\in \futon)$ and review some results from 
\cite{6-persons}, \cite{UR} in the case of Grassmann mnnifolds. 
As a whole introduction in this section \cite{MN} is recommended.

The Grassmann manifold is the set of complex vector spaces 
defined by
\begin{equation}
G_{m,n}(\fukuso)=\{{\cal V}\subset \fukuso^{m+n}\ |\ 
\mbox{dim}_{\fukuso}{\cal V}=m\}.
\end{equation}
Then it is well--known that
\begin{equation}
G_{m,n}(\fukuso)\cong U(m+n)/U(m)\times U(n)
\end{equation}
and moreover
\begin{equation}
U(m+n)/U(m)\times U(n)\cong GL(m+n)/B_{+}
\end{equation}
where $B_{+}$ is the (upper) Borel subgroup of $GL(m+n)$ given by
\[
B_{+}=
\left\{
\left(
\begin{array}{cc}
*_{1}   & * \\
{\bf 0} & *_{2}
\end{array}
\right)
\in GL(m+n)\ |\ 
*_{1}\in GL(m;\fukuso),\ *_{2}\in GL(n;\fukuso)
\right\}.
\]

In order to obtain the element of $U(m+n)/U(m)\times U(n)$ 
from an element in $GL(m+n)/B_{+}$ it is convenient to use 
the orthonormalization (method) by Gram--Schmidt. 

For the matrix
\begin{equation}
G\equiv
\left(
\begin{array}{cc}
E_{m}   & {\bf 0} \\
Z       & E_{n}
\end{array}
\right)
\ \in\ GL(m+n)/B_{+}
\end{equation}
where $Z\in M(n,m;\fukuso)$ we set
\[
V_{1}
=
\left(
\begin{array}{c}
E_{m} \\
Z
\end{array}
\right),\quad
V_{2}
=
\left(
\begin{array}{c}
{\bf 0} \\
E_{n}   
\end{array}
\right).
\]

For $\{V_{1},V_{2}\}$ the Gramm--Schmidt orthonormalization 
in the matrix form reads
\begin{eqnarray*}
\hat{V}_{1}&=&V_{1}(V_{1}^{\dagger}V_{1})^{-1/2}
\ \Longrightarrow\ 
P_{1}=\hat{V}_{1}{\hat{V}_{1}}^{\dagger}\ :\ \mbox{projection} \\
\tilde{V}_{2}&=&(E_{m+n}-P_{1})V_{2},\quad 
\hat{V}_{2}=\tilde{V}_{2}(\tilde{V}_{2}^{\dagger}\tilde{V}_{2})^{-1/2}.
\end{eqnarray*}
Explicitly,
\[
\hat{V}_{1}=
\left(
\begin{array}{c}
E_{m} \\
Z   
\end{array}
\right) L_{Z}^{-1/2},
\quad
\hat{V}_{2}=
\left(
\begin{array}{c}
-Z^{\dagger} \\
E_{n}  
\end{array}
\right) M_{Z}^{-1/2}
\]
where 
\[
L_{Z}=E_{m}+Z^{\dagger}Z,\quad M_{Z}=E_{n}+ZZ^{\dagger}.
\]
Therefore we obtain the unitary matrix
\begin{equation}
\hat{V}
=(\hat{V}_{1},\hat{V}_{2})
=
\left(
\begin{array}{cc}
E_{m} & -Z^{\dagger} \\
Z     & E_{n}
\end{array}
\right)
\left(
\begin{array}{cc}
L_{Z}^{-1/2} &                \\
              & M_{Z}^{-1/2}
\end{array}
\right)\ \in\ U(m+n).
\end{equation}

\vspace{5mm}
Next, if we consider a transformation
\[
G\ \longrightarrow\ 
G
\left(
\begin{array}{cc}
U_{1} &       \\
      & U_{2} 
\end{array}
\right)
\quad
\mbox{where}
\quad
\left(
\begin{array}{cc}
U_{1} &       \\
      & U_{2} 
\end{array}
\right)\ \in\ U(m)\times U(n)
\]
then the resultant unitary matrix $\hat{V}$ also transforms like
\[
\hat{V}\ \longrightarrow\ 
\hat{V}
\left(
\begin{array}{cc}
U_{1} &       \\
      & U_{2} 
\end{array}
\right).
\]
As a result, the procedure to obtain unitary matrices is {\bf covariant} 
under the subgroup $U(m)\times U(n)$, so we can consider 
$\hat{V}\in G_{m,n}$.

\vspace{5mm}
A comment is in order.\ As another definition of the Grassmann manifold 
the following one in terms of projections is also well--known. See \cite{HO1} 
and \cite{KF1} as an elementery introduction and \cite{FKS} as an 
advanced one.
\[
\{P\in M(m+n;\fukuso)\ |\ P^{2}=P,\ P^{\dagger}=P,\ 
\mbox{tr}P=m\}\cong U(m+n)/U(m)\times U(n).
\]
The correspondence ($\hat{V}\ \longrightarrow\ P$) is as follows :
\[
P
=
\hat{V}
\left(
\begin{array}{cc}
E_{m} &         \\
      & {\bf 0} 
\end{array}
\right)
\hat{V}^{-1}
=
\left(
\begin{array}{cc}
E_{m} & -Z^{\dagger} \\
Z     & E_{n}
\end{array}
\right)
\left(
\begin{array}{cc}
E_{m} &         \\
      & {\bf 0} 
\end{array}
\right)
\left(
\begin{array}{cc}
E_{m} & -Z^{\dagger} \\
Z     & E_{n}
\end{array}
\right)^{-1}.
\]

\vspace{5mm}
Next, we treat the time--dependent Hamiltonian (which is 
not so artificial) like
\begin{equation}
H=H(t)
=
\left(
\begin{array}{cc}
H_{1}(t) & V^{\dagger}(t) \\
V(t)     & H_{2}(t)
\end{array}
\right)
\ \in\ H(m+n;\fukuso).
\end{equation}
Then the (reduced) evolution equation
\begin{equation}
\label{eq:reduced ev-equation}
i\frac{d}{dt}{\hat{V}}=H(t)\hat{V}
\quad \mbox{where}\quad 
\hat{V}=\hat{V}(t)\ \Longleftrightarrow\ Z=Z(t)
\end{equation}
reduces to a matrix Riccati equation
\begin{equation}
i\dot{Z}=V+H_{2}Z-ZH_{1}-ZV^{\dagger}Z,
\end{equation}
see \cite{6-persons}, \cite{UR}. 
Note that we ignored the time evolution on $U_{1}$ and 
$U_{2}$, which is out of interest at the present time. 

\vspace{5mm}
Especially, in the case of $m=n=1$ 
\begin{eqnarray}
H
&=&
\left(
\begin{array}{cc}
h_{1}(t) & \bar{v}(t) \\
v(t)     & h_{2}(t)
\end{array}
\right) \\
\hat{V}
&=&
\frac{1}{\sqrt{1+|z(t)|^{2}}}
\left(
\begin{array}{cc}
1    & -\bar{z}(t) \\
z(t) & 1
\end{array}
\right)
\end{eqnarray}
the (matrix) Riccati equation becomes 
\begin{equation}
i\dot{z}=v+(h_{2}-h_{1})z-\bar{v}z^{2}.
\end{equation}

It is very interesting to note that this equation satisfies  
a mysterious formula called a {\bf nonlinear superposition}. 
See for example \cite{CL} and its references. As a whole 
introduction to the Riccati equation see \cite{WR}. 

Namely, let $z_{1},\ z_{2},\ z_{3}$ be three different 
solutions and $z$ be any solution. For the cross--ratio 
defined by
\begin{equation}
(z,z_{1},z_{2},z_{3})\equiv 
\frac{z-z_{1}}{z-z_{3}}\div \frac{z_{2}-z_{1}}{z_{2}-z_{3}}
\end{equation}
it is easy to see
\[
i\frac{d}{dt}(z,z_{1},z_{2},z_{3})=0
\ \Longrightarrow\ (z,z_{1},z_{2},z_{3})=k
\]
with constant $k\ (\in \fukuso)$. 
From this we can express $z$ in terms of $k$ and three solutions 
$z_{1},\ z_{2},\ z_{3}$ like
\begin{equation}
\label{eq:nonlinear superposition}
z=\frac{kz_{3}(z_{2}-z_{1})-z_{1}(z_{2}-z_{3})}
       {k(z_{2}-z_{1})-(z_{2}-z_{3})}.
\end{equation}

This is called the nonlinear superposition formula for the 
Riccati equation. Concerning this formula we have the 
following

\vspace{5mm}
\noindent
{\bf Problem}\ \ Can this nonlinear superposition formula 
be derived (or reduced) from usual one in the linear equation 
(\ref{eq:reduced ev-equation}) ?

\section{Reduced Dynamics on Flag Manifolds}

In this section we generalize the results in the preceding 
section based on Grassmann manifolds to Flag manifolds 
of the second type (in our terminology). 
See \cite{Pi} and \cite{DJ} as a good introduction.

For $l+m+n=N\ (l, m, n\in \futon)$ the flag manifold 
of the second type is the sequence of complex vector spaces 
defined by
\begin{equation}
F_{l,m,n}(\fukuso)=\{{\cal V}\subset {\cal W}\subset \fukuso^{l+m+n}\ 
|\ \mbox{dim}_{\fukuso}{\cal V}=l,\ \mbox{dim}_{\fukuso}{\cal W}=l+m\}.
\end{equation}

Then it is well--known that
\begin{equation}
F_{l,m,n}(\fukuso)\cong U(l+m+n)/U(l)\times U(m)\times U(n)
\end{equation}
and moreover
\begin{equation}
U(l+m+n)/U(l)\times U(m)\times U(n)\cong GL(l+m+n)/B_{+}
\end{equation}
where $B_{+}$ is the Borel subgroup given by
\[
B_{+}=
\left\{
\left(
\begin{array}{ccc}
*_{1}   & *       & *     \\
{\bf 0} & *_{2}   & *     \\
{\bf 0} & {\bf 0} & *_{3} \\
\end{array}
\right)
\in GL(l+m+n)\ |\ 
*_{1}\in GL(l;\fukuso),\ *_{2}\in GL(m;\fukuso),\ 
*_{3}\in GL(n;\fukuso)
\right\}.
\]

Similarly in the preceding section we consider the matrix
\begin{equation}
F\equiv
\left(
\begin{array}{ccc}
E_{l} & {\bf 0} & {\bf 0} \\
X     & E_{m}   & {\bf 0} \\
Y     & Z       & E_{n}
\end{array}
\right)
\ \in\ GL(l+m+n)/B_{+}
\end{equation}
and set
\[
V_{1}
=
\left(
\begin{array}{c}
E_{l} \\
X     \\
Y
\end{array}
\right),\quad
V_{2}
=
\left(
\begin{array}{c}
{\bf 0} \\
E_{m}   \\
Z
\end{array}
\right),\quad
V_{3}
=
\left(
\begin{array}{c}
{\bf 0} \\
{\bf 0} \\
E_{n}
\end{array}
\right).
\]

\vspace{5mm}
For $\{V_{1},V_{2},V_{3}\}$ we perform the Gramm--Schmidt orthogonalization 
in the matrix form like
\begin{eqnarray*}
\hat{V}_{1}&=&V_{1}(V_{1}^{\dagger}V_{1})^{-1/2}
\ \Longrightarrow\ 
P_{1}=\hat{V}_{1}{\hat{V}_{1}}^{\dagger}\ :\ \mbox{projection}  \\
\tilde{V}_{2}&=&(E_{l+m+n}-P_{1})V_{2},\quad 
\hat{V}_{2}=\tilde{V}_{2}(\tilde{V}_{2}^{\dagger}\tilde{V}_{2})^{-1/2}
\ \Longrightarrow\ 
P_{2}=\hat{V}_{2}{\hat{V}_{2}}^{\dagger}\ :\ \mbox{projection}  \\
\tilde{V}_{3}&=&
(E_{l+m+n}-P_{1}-P_{2})V_{3}=(E_{l+m+n}-P_{1})(E_{l+m+n}-P_{2})V_{3}
,\quad 
\hat{V}_{3}=\tilde{V}_{3}(\tilde{V}_{3}^{\dagger}\tilde{V}_{3})^{-1/2}.
\end{eqnarray*}

\noindent
Explicitly,
\begin{eqnarray*}
\hat{V}_{1}&=&
\left(
\begin{array}{c}
E_{l} \\
X     \\
Y
\end{array}
\right) \Lambda^{-1/2},  \\
\hat{V}_{2}&=&
\left(
\begin{array}{c}
-\Lambda^{-1}\delta^{\dagger}       \\
E_{m}-X\Lambda^{-1}\delta^{\dagger} \\
Z-Y\Lambda^{-1}\delta^{\dagger}
\end{array}
\right) 
\left(L-\delta\Lambda^{-1}\delta^{\dagger}\right)^{-1/2},  \\
\hat{V}_{3}&=&
\left(
\begin{array}{c}
-\Lambda^{-1}Y^{\dagger}+\Lambda^{-1}\delta^{\dagger}
\left(L-\delta\Lambda^{-1}\delta^{\dagger}\right)^{-1}
(Z-Y\Lambda^{-1}\delta^{\dagger})^{\dagger}                 \\
-X\Lambda^{-1}Y^{\dagger}-(E_{m}-X\Lambda^{-1}\delta^{\dagger})
\left(L-\delta\Lambda^{-1}\delta^{\dagger}\right)^{-1}
(Z-Y\Lambda^{-1}\delta^{\dagger})^{\dagger}                 \\
E_{n}-Y\Lambda^{-1}Y^{\dagger}-(Z-Y\Lambda^{-1}\delta^{\dagger})
\left(L-\delta\Lambda^{-1}\delta^{\dagger}\right)^{-1}
(Z-Y\Lambda^{-1}\delta^{\dagger})^{\dagger}                 
\end{array}
\right) \times                                              \\
&&
\ \ \left(
E_{n}-Y\Lambda^{-1}Y^{\dagger}-
(Z-Y\Lambda^{-1}\delta^{\dagger})
\left(L-\delta\Lambda^{-1}\delta^{\dagger}\right)^{-1}
(Z-Y\Lambda^{-1}\delta^{\dagger})^{\dagger}
\right)^{-1/2}
\end{eqnarray*}
where
\[
\Lambda=E_{l}+X^{\dagger}X+Y^{\dagger}Y,\quad
L=E_{m}+Z^{\dagger}Z,\quad
\delta=X+Z^{\dagger}Y.
\]
Therefore we obtain the unitary matrix
\begin{equation}
\hat{V}=(\hat{V}_{1},\hat{V}_{2},\hat{V}_{3})\ \in\ U(l+m+n).
\end{equation}

Here we list a decomposition of V
\begin{small}
\begin{eqnarray*}
&{}&
\hat{V}=  \\
&{}&
\left(
\begin{array}{ccc}
E_{l} & {\bf 0} & {\bf 0} \\
X     & E_{m}   & {\bf 0} \\
Y     & Z       & E_{n}
\end{array}
\right)\times  \\
&{}&
\left(
\begin{array}{ccc}
E_{l} & -\Lambda^{-1}\delta^{\dagger} & 
-\Lambda^{-1}Y^{\dagger}+\Lambda^{-1}\delta^{\dagger}
(L-\delta\Lambda^{-1}\delta^{\dagger})^{-1}
(Z-Y\Lambda^{-1}\delta^{\dagger})^{\dagger}             \\
{\bf 0} & E_{m} & 
-(L-\delta\Lambda^{-1}\delta^{\dagger})^{-1}
(Z-Y\Lambda^{-1}\delta^{\dagger})^{\dagger}             \\
{\bf 0} & {\bf 0} & E_{n}
\end{array}
\right)\times  \\
&{}&
\left(
\begin{array}{ccc}
\Lambda^{-1/2} &                                      &  \\
      & (L-\delta\Lambda^{-1}\delta^{\dagger})^{-1/2} &  \\
      &        & 
(E_{n}-Y\Lambda^{-1}Y^{\dagger}-
(Z-Y\Lambda^{-1}\delta^{\dagger})
\left(L-\delta\Lambda^{-1}\delta^{\dagger}\right)^{-1}
(Z-Y\Lambda^{-1}\delta^{\dagger})^{\dagger}
)^{-1/2}
\end{array}
\right).
\end{eqnarray*}
\end{small}

\noindent
Compare $\hat{V}$ with the corresponding unitary matrix in \cite{Pi} 
and \cite{DJ} where $l=m=n=1$.

\vspace{5mm}
If we consider a transformation
\[
F\ \longrightarrow\ 
F
\left(
\begin{array}{ccc}
U_{1} &       &       \\
      & U_{2} &       \\
      &       & U_{3}
\end{array}
\right)
\quad \mbox{where}\quad
\left(
\begin{array}{ccc}
U_{1} &       &       \\
      & U_{2} &       \\
      &       & U_{3}
\end{array}
\right)
\in U(l)\times U(m)\times U(n)
\]
then the resultant unitary matrix $\hat{V}$ also transforms like
\[
\hat{V}\ \longrightarrow\ 
\hat{V}
\left(
\begin{array}{ccc}
U_{1} &       &       \\
      & U_{2} &       \\
      &       & U_{3}
\end{array}
\right).
\]
As a result, the procedure to obtain unitary matrices is covariant 
under the subgroup $U(l)\times U(m)\times U(n)$, so we can consider 
$\hat{V}\in F_{l,m,n}$.

\vspace{5mm}
A comment is in order.\ As another definition of the flag manifold 
the following one in terms of projections is also well--known.
\begin{eqnarray*}
&&\{(P,Q):\mbox{a pair of projections in}\ M(l+m+n;\fukuso)\ |\ 
\mbox{tr}P=l,\ \mbox{tr}Q=l+m,\ PQ=P\} \\
&&\cong U(l+m+n)/U(l)\times U(m)\times U(n).
\end{eqnarray*}
The correspondence ($\hat{V}\ \longrightarrow\ (P,Q)$) is as follows :
\begin{eqnarray*}
P
&=&
\hat{V}
\left(
\begin{array}{ccc}
E_{l} &         &          \\
      & {\bf 0} &          \\
      &         & {\bf 0}
\end{array}
\right)
\hat{V}^{-1}
=
W
\left(
\begin{array}{ccc}
E_{l} &         &          \\
      & {\bf 0} &          \\
      &         & {\bf 0}
\end{array}
\right)
W^{-1},  \\
Q
&=&
\hat{V}
\left(
\begin{array}{ccc}
E_{l} &       &          \\
      & E_{m} &          \\
      &       & {\bf 0}
\end{array}
\right)
\hat{V}^{-1}
=
W
\left(
\begin{array}{ccc}
E_{l} &       &          \\
      & E_{m} &          \\
      &       & {\bf 0}
\end{array}
\right)
W^{-1}
\end{eqnarray*}
where
\[
W=
\left(
\begin{array}{ccc}
E_{l} & -\Lambda^{-1}\delta^{\dagger}       & 
-\Lambda^{-1}Y^{\dagger}+\Lambda^{-1}\delta^{\dagger}
\left(L-\delta\Lambda^{-1}\delta^{\dagger}\right)^{-1}
(Z-Y\Lambda^{-1}\delta^{\dagger})^{\dagger}                    \\
X     & E_{m}-X\Lambda^{-1}\delta^{\dagger} & 
-X\Lambda^{-1}Y^{\dagger}-(E_{m}-X\Lambda^{-1}\delta^{\dagger})
\left(L-\delta\Lambda^{-1}\delta^{\dagger}\right)^{-1}
(Z-Y\Lambda^{-1}\delta^{\dagger})^{\dagger}                    \\
Y     & Z-Y\Lambda^{-1}\delta^{\dagger}     & 
E_{n}-Y\Lambda^{-1}Y^{\dagger}-(Z-Y\Lambda^{-1}\delta^{\dagger})
\left(L-\delta\Lambda^{-1}\delta^{\dagger}\right)^{-1}
(Z-Y\Lambda^{-1}\delta^{\dagger})^{\dagger} 
\end{array}
\right).
\]

\vspace{5mm}
Next, we treat the time--dependent Hamiltonian like
\begin{equation}
H=H(t)
=
\left(
\begin{array}{ccc}
H_{1}(t) & V_{1}^{\dagger}(t) & V_{2}^{\dagger}(t) \\
V_{1}(t) & H_{2}(t) & V_{3}^{\dagger}(t)           \\
V_{2}(t) & V_{3}(t) & H_{3}(t)
\end{array}
\right).
\end{equation}
Then the (reduced) evolution equation
\begin{equation}
i\frac{d}{dt}\hat{V}=H(t)\hat{V}
\quad \mbox{where}\quad \hat{V}=\hat{V}(t)
\ \Longleftrightarrow\ 
X=X(t),\ Y=Y(t),\ Z=Z(t)
\end{equation}
gives a set of matrix Riccati equations interecting with 
one another
\begin{eqnarray}
i\dot{X}&=&V_{1}+H_{2}X-XH_{1}-XV_{1}^{\dagger}X+
          V_{3}^{\dagger}Y-XV_{2}^{\dagger}Y, \nonumber \\
i\dot{Y}&=&V_{2}+H_{3}Y-YH_{1}-YV_{2}^{\dagger}Y+
          V_{3}X-YV_{1}^{\dagger}X, \\
i\dot{Z}&=&V_{3}+H_{3}Z-ZH_{2}-ZV_{3}^{\dagger}Z+
          (ZX-Y)(V_{1}^{\dagger}+V_{2}^{\dagger}Z). \nonumber
\end{eqnarray}
This is our main result in the paper.

\vspace{5mm}
Especially, in the case of $l=m=n=1$ 
\begin{eqnarray}
H&=&
\left(
\begin{array}{ccc}
h_{1}(t) & \bar{v}_{1}(t) & \bar{v}_{2}(t) \\
v_{1}(t) & h_{2}(t)       & \bar{v}_{3}(t) \\
v_{2}(t) & v_{3}(t)       & h_{3}(t)
\end{array}
\right) \\
\hat{V}&=&
\left(
\begin{array}{ccc}
\frac{1}{\sqrt{\Delta_{1}}} & 
\frac{-(\bar{x}+\bar{y}z)}{\sqrt{\Delta_{1}\Delta_{2}}} &
\frac{\bar{x}\bar{z}-\bar{y}}{\sqrt{\Delta_{2}}}            \\
\frac{x}{\sqrt{\Delta_{1}}} & 
\frac{1-(xz-y)\bar{y}}{\sqrt{\Delta_{1}\Delta_{2}}} &
\frac{-\bar{z}}{\sqrt{\Delta_{2}}}                          \\
\frac{y}{\sqrt{\Delta_{1}}} &
\frac{z+\bar{x}(xz-y)}{\sqrt{\Delta_{1}\Delta_{2}}} &
\frac{1}{\sqrt{\Delta_{2}}}
\end{array}
\right)
\end{eqnarray}
where
\[
\Delta_{1}=1+|x|^{2}+|y|^{2},\quad
\Delta_{2}=1+|z|^{2}+|xz-y|^{2}
\]
the interacting Riccati equations become
\begin{eqnarray}
i\dot{x}&=&v_{1}+(h_{2}-h_{1})x-\bar{v}_{1}x^{2}+
           \bar{v}_{3}y-\bar{v}_{2}xy,          \nonumber \\
i\dot{y}&=&v_{2}+(h_{3}-h_{1})y-\bar{v}_{2}y^{2}+
           v_{3}x-\bar{v}_{1}xy,                          \\
i\dot{z}&=&v_{3}+(h_{3}-h_{2})z-\bar{v}_{3}z^{2}+
           (xz-y)(\bar{v}_{1}+\bar{v}_{2}z).    \nonumber
\end{eqnarray}
Concerning these equations, our interest is the following

\noindent
{\bf Problem}\ \ In this system is there a nonlinear 
superposition formula like (\ref{eq:nonlinear superposition}) ?

This is an important problem on nonlinear superposition. However, 
we cannot find such a formula at the present time, so we leave it 
to readers as a challenging problem.

\section{Discussion}

In this paper we considered the reduced dynamics from 
the unitary group $U(N)$ to flag manifolds of the second type 
and derived a set of differential equations of matrix Riccati 
types interacting with one another. 
 
We also presented an important problem in the special case 
($l=m=n=1$) on generalization of the nonlinear superposition 
formula that the Riccati equation satisfies.

In last, let us make a comment. 
We can generalize our construction to flag manifolds 
of the third type (in our terminology)
\[
F_{k,l,m,n}\cong U(k+l+m+n)/U(k)\times U(l)\times U(m)\times U(n)
\]
where
\[
F_{k,l,m,n}(\fukuso)=\{{\cal U}\subset {\cal V}\subset {\cal W}\subset 
\fukuso^{k+l+m+n}\ 
|\ \mbox{dim}_{\fukuso}{\cal U}=k,\ \mbox{dim}_{\fukuso}{\cal V}=k+l,\ 
\mbox{dim}_{\fukuso}{\cal W}=k+l+m\}.
\]

In fact, starting from the matrix
\[
F\equiv
\left(
\begin{array}{cccc}
E_{k} & {\bf 0} & {\bf 0} & {\bf 0} \\
K     & E_{l}   & {\bf 0} & {\bf 0} \\
L     & X       & E_{m}   & {\bf 0} \\
M     & Y       & Z       & E_{n}
\end{array}
\right)
\ \in\ GL(k+l+m+n)/B_{+}
\]
we can trace the same process as the text. Explicit calculations done 
are of course very hard, see \cite{HO2}. 

When $k=l=m=n=1$ some explicit calculations have been done by 
Picken \cite{Pi}.



\begin{thebibliography}{99}
%
\bibitem{6-persons} S. Chaturvedi, E. Ercolessi, G. Marmo, 
G. Morandi, N. Mukunda and R. Simon :  
\newblock Ray space `Riccati' evolution and geometric phases for 
$N$--level quantum systems, 
\newblock Pramana J. Phys. 69(2007) 317
\newblock arXiv:0706.0964 [quant-ph].
%
\bibitem{UR}D. B. Uskov and A. R. P. Rau : 
\newblock Geometric phase for SU(N) through fiber bundles 
and unitary integration, 
\newblock Phys.Rev.A 74, 030304(R) (2006), 
\newblock quant-ph/0511192. 
%
\bibitem{CGM}J. F. Carine{\~n}a, J. Grabowski and G. Marmo : 
\newblock Lie--Scheffers Systems : A Geometric Approach, 
\newblock 2000, Bibliopolis, Naples.
%
\bibitem{MN}M. Nakahara : 
\newblock GEOMETRY, TOPOLOGY AND PHYSICS (Second Edition), 
\newblock 2003, Taylor $\&$ Francis.
%
\bibitem{HO1}H. Oike : 
\newblock Introduction to Grassmann Manifolds (in Japanese), 
\newblock 1978, Lecture Note (Yamagata University).
%
\bibitem{KF1}K. Fujii : 
\newblock Introduction to Grassmann Manifolds and Quantum Computation, 
\newblock J. Applied Math, 2(2002), 371, 
\newblock quant-ph/0103011. 
%
\bibitem{FKS}K. Fujii, T. Kashiwa, S. Sakoda : 
\newblock Coherent states over Grassmann manifolds and the WKB exactness
in path integral,
\newblock J. Math. Phys., 37(1996), 567.
%
\bibitem{CL}J. F. Carine{\~n}a and J. de Lucas : 
\newblock A nonlinear superposition rule for solutions of the 
Milne--Pinney equation, 
\newblock arXiv:0807.0370 [math-ph].
%
\bibitem{WR}W. T. Reid : 
\newblock Riccati Differential Equations, 
\newblock 1972, Academic Press.
%
\bibitem{Pi}R. F. Picken : 
\newblock The Duistermaat--Heckman Integration Formula On Flag Manifolds, 
\newblock J. Math. Phys. 31(1990), 616.
%
\bibitem{DJ}M. Daoud and A. Jellal : 
\newblock Quantum Hall Effect on the Flag Manifold F$_{2}$, 
\newblock Int. J. Mod. Phys. A, 
\newblock hep-th/0610157. 
%
\bibitem{HO2}H. Oike : 
\newblock in preparation.
\end{thebibliography}
\end{document}